# Corpus-based Web Document Summarization using Statistical and Linguistic Approach

Rushdi Shams[1], M.M.A. Hashem[2], Afrina Hossain[3], Suraiya Rumana Akter[4], and Monika Gope[5]
Department of Computer Science and Engineering
Khulna University of Engineering & Technology (KUET)
Khulna 9203, Bangladesh
{rushdecoder[1], mma_hashem[2], afrina_cse[3], rumana_kuet[4], gopenath14[5]}@yahoo.com

*Abstract*— Single document summarization generates summary by extracting the representative sentences from the document. In this paper, we presented a novel technique for summarization of domain-specific text from a single web document that uses statistical and linguistic analysis on the text in a reference corpus and the web document. The proposed summarizer uses the combinational function of Sentence Weight ($SW$) and Subject Weight ($SuW$) to determine the rank of a sentence, where $SW$ is the function of number of terms ($t_n$) and number of words ($w_n$) in a sentence, and term frequency ($t_f$) in the corpus and $SuW$ is the function of $t_n$ and $w_n$ in a subject, and $t_f$ in the corpus. 30 percent of the ranked sentences are considered to be the summary of the web document. We generated three web document summaries using our technique and compared each of them with the summaries developed manually from 16 different human subjects. Results showed that 68 percent of the summaries produced by our approach satisfy the manual summaries.

## I. INTRODUCTION

The number of pages available on the Internet almost doubles every year [1]. In July 2009, the number of hosts advertised in the DNS is 681,064,561 [2]. This bulk forces the search engines to provide numerous web pages for a single search. To find the desired information, user often has to browse hundreds of pages where only a few of them are relevant. Most of the users also have limited knowledge regarding the relevance and appropriateness of the information in the pages because of absence of contextual and discourse awareness in today's web. Therefore, summarizing all the information with contextual and discourse awareness is helpful for the user to find out relevant and appropriate information from the web.

If the user searches the web with a keyword *resistance*, the web may return him the pages containing the information of *resistance of electricity* and *resistance of body against diseases*. The key missing here is the relevance and appropriateness of text according to the domain, context and discourse of the keyword. Therefore, successful web text summarization depends on the measurement of relevance between the text in the web and a reference that can be a corpus- structured and representative collection of text.

This paper presents an approach to summarize domain-specific text from single web document using both linguistic and statistical methods. To achieve this, we introduced two novel features- Sentence Weight ($SW$) and Subject Weight ($SuW$) to rank sentences and used a representative domain-specific corpus for the domain DC electrical circuits [3] [10] [13]. Here, $SW$ is the function of number of terms ($t_n$) and number of words ($w_n$) in a sentence, and term frequency ($t_f$) in the corpus and $SuW$ is the function of $t_n$ and $w_n$ in a subject, and $t_f$ in the corpus. We considered 30 percent of the ranked sentences as the summary of the document. Experimental results showed that 68 percent of the summaries produced by our approach satisfy the manual summaries produced by 16 different human subjects.

The organization of the paper is as follows. Section II discusses the leading techniques of document summarization. In Section III, we discuss the proposed approach of text summarization. Experimental results and related discussions are depicted in Section IV. Section V concludes the paper.

## II. RELATED WORK

In this section, we discuss about SweSum- a summarization tool from the Royal Institute of Technology, Sweden. We also refer MEAD- a public domain multi-lingual multi-document summarization system developed by the research group of Dragomir Radev. Lastly, we discuss LEMUR- a summarizer toolkit that provides the summary with its own search engine.

### A. SweSum

SweSum [4], an online summarizer, was first constructed by Hercules Dalianis in 1999, and further developed by Martin Hassel. It is a traditional extraction-based domain-specific text summarizer that works on sentences from news text using HTML tags. It utilizes a lexicon for mapping inflected forms of content words for each language. For topic identification, SweSum applies the hypothesis, where the high-frequent content words are keys to the topic of the text. In news paper text, the most relevant information is often presented at the top. Frequencies are modified by a set of heuristics, e.g., the position of the sentence in the text and its formatting.

Sentences that contain keywords are scored high. A keyword is an open class word with a high Term Frequency. Sentences containing numerical data are also considered carrying important information. These parameters are put into a combination function with modifiable weights to obtain the total score of each sentence. For Swedish, SweSum also features anaphora resolution as well as named entity tagging [6]. Complete user dependency and absence of generic summary makes it difficult for inexpert user to set the parameter of the SweSum.

### B. Mead

Mead is a centroid-based extractive domain-specific summarizer that scores sentences based on sentence-level and inter-sentence features which indicate the quality of the sentence as a summary sentence [7]. It chooses the top-ranked sentences for inclusion in the output summary. MEAD extractive summaries score sentences according to certain sentence features- centroid position and length. It only works with the news text, not with the web pages- which are significantly different in nature, structure and presentation.

### C. Lemur

Lemur [10] is a toolkit which not only searches the web but also makes summary of both single and multi-documents. It utilizes ad hoc retrieval, TFIDF (vector model), Okapi (probabilistic model) for multi-document, and structured query language as relevance feedback. Lemur takes newswire document files and breaks each one into individual "documents" based on the <DOC> formatting tags in the files. Also, Lemur provides a standard tokenizer (e.g., a parser) that has options for stemming and stop-words.

### III. PROPOSED TEXT SUMMARIZATION APPROACH

Summary is mainly concerned with judging the importance or the indicative power of each sentence in a given document [9]. There are two common approaches used in summarization- the statistical approach and the linguistic approach. Statistical approaches derive weights of key terms and determine the importance of sentence by the total weight contained by the sentence, whereas linguistics-based approaches identify term relationships in the document through Part of Speech (POS) tagging, grammar analysis, thesaurus usage, and extract meaningful sentences. Statistical approaches maybe efficient in computation but linguistic approaches look into term semantics, which may yield better summary results [11] - in our proposed summarization, we used both of these approaches. We used a representative multimodal corpus for the domain *DC Electrical Circuits* that contains over 1,000 sentences from 144 web resources. We selected three web documents containing text for the domain and named them as Document 1 [14], Document 2 [15], and Document 3 [16].

The proposed summarization technique works in three steps. First, it selects web documents that are resourceful with respect to the domain. Second, the summarizer extracts text from these web documents. Lastly, it summarizes the extracted text.

### A. Identifying the Resourceful Document

We calculate the mean of each document to determine their resourcefulness. Thereafter, the mean of each document is compared with the mean of the corpus. In this case, the mean,

$$\mu = \frac{\sum_{i=1}^{N} S_{1_i}}{N} \qquad (1)$$

Where, $S_1$= Sentence weight and $N$ = Number of sentence in the document

Both the corpus and the documents use (1) to calculate their respective means. On the other hand, sentence weight is calculated using following equation-

$$S_1 = P(a\ sentence\ to\ be\ representative) \times \sum term\ frequency\ in\ sentence$$

$$= \frac{t_n}{w_n} \times \sum_{k=1}^{t_n} tf_k \qquad (2)$$

Where, $t_n$ = Number of terms (Noun) in a sentence, $w_n$ = Number of words in the sentence, $tf_k$ =Term frequency of $k^{th}$ term in the sentence.

$$P(a\ sentence\ to\ be\ representative) = \frac{Number\ of\ term\ frequencies\ in\ a\ sentence,\ t_n}{Number\ of\ words\ in\ a\ sentence, w_n}$$

The sentence weight is equal to the summation of term frequency multiplied by probability of the sentence to be representative. This probability is a ratio of $t_n/w_n$ in order to get the effect of the length of the sentence on $\sum tf$. Document with mean distant from the mean of the corpus in a positive direction is more informative in the domain and is chosen for summarization.

From Fig. 1, we see that documents 1, 2, 3, 4 and 5 have mean 6.12, 4.92, 10.14, 1.99 and 4.88, respectively where corpus is denoted as document 1.

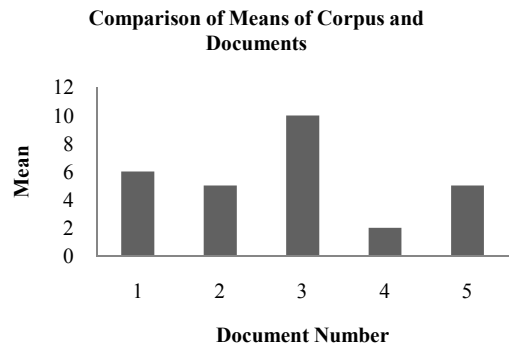

Figure 1.  Comparison of mean of corpus and documents

From the documents, document 3 is more informative than the others as the distance of its mean from the mean of the corpus is the highest.

## B. Extracting Text from Web Document

Web documents have defined structures and they consist of some sections and subsections like abstract, introduction, application, types, and advantages. We converted this structured text into flat text where we preserved the paragraph only. First, we selected an HTML document and removed all of its tags (except <p>) to collect flat text. Thereafter, we collected text that is separated by the tag <p>- means the text is divided into paragraphs. Within a paragraph, sentences are separated by period ".".

## C. Summarization of Text

The proposed approach uses natural language processing techniques for summarization purposes as well as the statistical methods on term frequency. Any text summary can be either query-relevant summary or generic summary. A query-relevant summary presents the contents of the document that are closely related to the initial search query. Creating a query-relevant summary is essentially a process of retrieving the query relevant sentences from the document. On the other hand, a generic summary provides an overall sense of the contents of documents and it should contain the main topics of the documents. Our proposed method utilizes both types of text summarization.

We considered four properties of a sentence- sentence weight ($SW$), subject weight ($SuW$), correlation between two sentences, and their positions. We considered another property called sentence rank- a combination of $SW$ and $SuW$.

From the corpus, we calculated Term Frequency ($t_f$) of the nouns. Table I shows the $t_f$ of the nouns that are present most frequently in the corpus.

TABLE I. TERM FREQUENCY IN THE CORPUS

| Term (noun) | Term Frequency | Term (noun) | Term Frequency |
|---|---|---|---|
| Current | 31.8584 | Electric | 3.5398 |
| Charge | 49.1150 | Ohm | 10.1769 |
| Circuit | 100.0 | Unit | 26.9911 |
| Voltage | 73.0088 | Series | 22.1238 |
| Power | 18.5840 | Law | 20.7964 |
| Resistance | 75.2212 | Wire | 29.6460 |
| Energy | 49.1150 | Battery | 20.7964 |

For example, we can consider the following sentence- "*The DC solution of an electric circuit is the solution where all voltages and currents are constant*",

The summation of term frequency,

$$\sum term\ frequency\ in\ sentence = \sum_{k=1}^{t_n} tf_k \quad (3)$$

= 8.40708+3.539823+100.0+1.3274336+73.00885+31.858408
= 218.14159

Here, Number of terms, $t_n$ = 6, Number of words in a sentence, $w_n$ = 17

From (2), we find the sentence weight,

$$S_1 = \frac{t_n}{w_n} \times \sum_{k=1}^{t_n} tf_k = \frac{6}{17} \times 218.12159 = 76.9841$$

Then, $S_1$ is normalized with the maximum weight of a sentence in the document in the following way-

$$SW = \frac{Sentence\ Weight,\ S_1}{Maximum\ Sentence\ Weight,\ \max(S_1)} \quad (4)$$

Geng et al. [13] proposed a summarization system which is based on the subject information from term co-occurrence graph and linkage information of different subjects. In this research, the term co-occurrence graph of the document is generated after term co-occurrences are calculated. Thereafter, the graph is sub-divided into many connected subjects- which are most significant linguistic information in a sentence. In our approach, the subject is determined from the sentence structure and then weight of each subject is summed up with the sentence weight.

In this regard, first, POS tagging is performed for each sentence. We used Stanford POS Tagger [17] to tag a sentence. For example, after tagging the sentence, we determined its sentence structure as-

The DC solution (NP) + of an electric circuit (PP) +is (VBZ) + the solution (NP) + where (WRB) + all voltages (NP) + and (CC) + currents (NNS) + are (VBP) + constant (VBN)

Here, the noun phrase (NP) is, NP → DT (JJ)* NN*, where NN is NNP or NNS and preposition Phrase is combination of PP → IN NP. The NNS, NNP, NP or NP+PP left to the verb are considered as the subject. In this sentence, the subject is - *The DC solution of an electric circuit*.

Then, subject weight is calculated from the $t_f$ of corpus as follows-

$$S_2 = \sum term\ frequency\ in\ subject = \sum_{k=1}^{t_n} tf_k \quad (5)$$

For the example, we found $S_2$ of the sentence is 108.8495.

Then, the weight of each subject is divided by the maximum subject weight.

$$SuW = \frac{Subject\ Weight,\ S_2}{Maximum\ Subject\ Weight,\ \max(S_2)} \quad (6)$$

The rank of a sentence is the combination of the $SW$ and $SuW$ for the generic summary.

We get the rank of a sentence from (4) and (6) as follows-

$$R = SW + SuW \qquad (7)$$

For the whole document and the corpus, our example sentence has *SW* of 41.76 and *SuW* of 63.7296 and the rank of the sentence, *R* is 105.48796. We ranked each sentence in every document using (7) and the sorted ranks of the sentences of Document 1 are depicted in Table II.

TABLE II.     SENTENCE RANKS AND RANK VALUES FOR DOCUMENT 1

| Rank | Rank Value | Rank | Rank Value |
|---|---|---|---|
| 1 | 191.46 | 17 | 34.66 |
| 2 | 150.24 | 18 | 34.42 |
| 3 | 142.11 | 19 | 27.70 |
| 4 | 137.26 | 20 | 24.80 |
| 5 | 76.10 | 21 | 23.73 |
| 6 | 70.21 | 22 | 19.42 |
| 7 | 69.68 | 23 | 17.23 |
| 8 | 66.65 | 24 | 15.52 |
| 9 | 66.22 | 25 | 15.01 |
| 10 | 63.84 | 26 | 10.27 |
| 11 | 59.25 | 27 | 9.63 |
| 12 | 49.25 | 28 | 9.37 |
| 13 | 46.06 | 29 | 5.52 |
| 14 | 45.51 | 30 | 4.41 |
| 15 | 44.39 | 31 | 0.73 |
| 16 | 42.69 | 32 | 0.49 |

We take 30 percent of every document as its summary. Document 1 contains 32 sentences and therefore 11 of these sentences will be included in its summary. So, the weight of the sentence with rank 11 becomes the lower bound to justify a sentence to be in summary, in this case which is 59.25. If the rank of any sentence is greater than 59.25, we selected that sentence for summary, otherwise not. Since, sentences are chosen sequentially and as we preserved the paragraphs by keeping the tag <p> in a sequential order, the information flow was completely preserved.

The correlation among sentences is very important for the summary as a sentence often refers to the previous or next sentence. In our approach, we only concentrated with the relation of a sentence with its previous sentence. We observed that the sentence starting with connectives like *such, beyond, although, however, moreover, also, this, these, those,* and *that* are related with preceding sentence. So, in such case, the sentence prior to the selected sentence for summary is considered to be included in the summary. If the rank of the referred sentence is greater than or equal to the 70 percent of the rank of the selected sentence, then it is included in the summary.

## IV. EXPERIMENTAL RESULTS

The comparison of our summary with the 16 manual summaries is shown in Fig. 2.

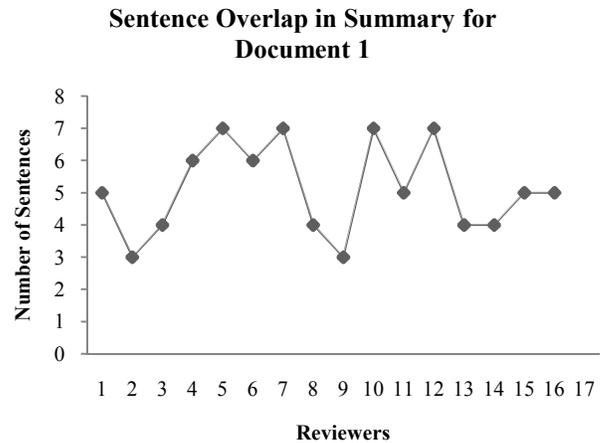

Figure 2.   Sentence overlap between the summary of document 1 and 16 human summaries

If we consider the summary of Document 1, from Fig. 2, we can see that the sentence overlap between the summary of our summarizer and the summary of reviewer 1 is five- means that the summary of the reviewer 1 contains five of the 11 sentences of our summary. The maximum number of sentence overlap for the document is seven (out of 11 sentences) and has been chosen by four reviewers.

Fig. 3 shows the comparison of our summary of Document 2 with the 16 manual summaries. The summary of Document 2 contains 12 sentences and the maximum number of sentence overlap between the summaries is nine.

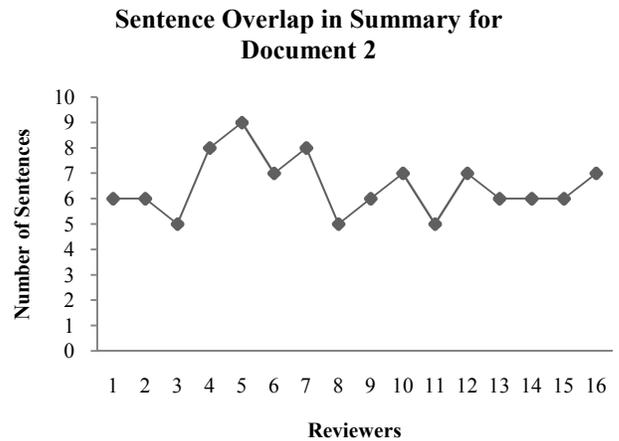

Figure 3.   Sentence overlap between the summary of document 2 and and 16 human summaries

Similarly, Fig. 4 shows that the summary of Document 3 contains six sentences and maximum number of sentence overlap between the summaries is five- means five sentences from the summary developed by our summarizer are also chosen by the human subjects in their summaries.

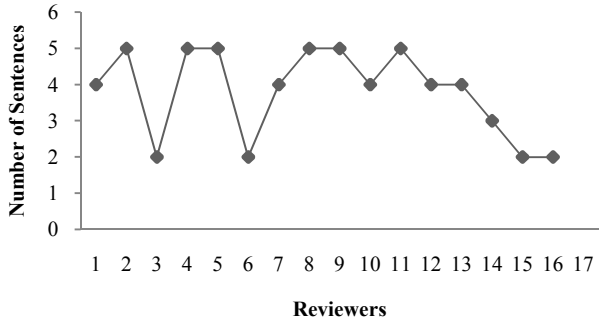

Figure 4.  Sentence overlap between the summary of document 3 and 16 human summaries

In Precision and Recall, generated summary is compared against an "ideal" reference summary to find sentence overlaps- which is the performance of the prior. To construct the reference summary, a group of human subjects are asked to extract sentences [12]. Then, the sentences chosen by a majority of humans are included in the reference summary. From the 16 manual summaries, we created a reference summary for each document and compared our summary with the reference summary. The comparison is described in Fig. 5-

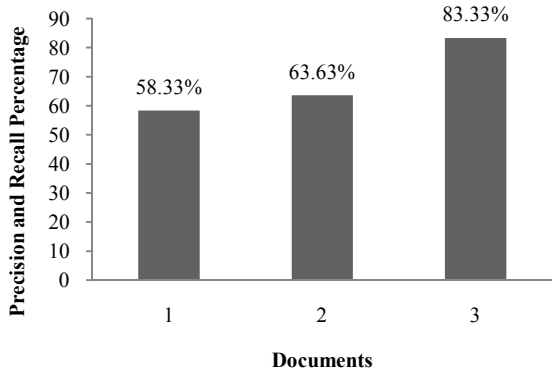

Figure 5.  Sentence overlaps between our summary and the reference summary using precision and recall

In Fig. 5, the horizontal line indicates the document numbers and the vertical line indicates the precision and recall percentage- sentences present both in the reference summary and the summary of our summarizer. This is the performance of our summarizer and it shows its efficiency in summarizing Document 1 is 58.33 percent, Document 2 is 63.63 percent and Document 3 is 83.33 percent, producing the average of 68.43 percent.

We calculate the Sample Standard Deviation (SSD) for both summaries to measure the deviation between them. For the most resourceful document, almost every line of it is important. From Table III, we see that Document 3, the most resourceful document, has the lowest difference between the Mean and SSD due to the similarity of sentence of the document and top-ranked sentences in the corpus. Therefore, we can evaluate the resourcefulness of a web document based on its Mean and SSD.

TABLE III.  RESULT OF MEAN AND STANDARD DEVIATION BETWEEN SUMMARIES

| Document | Mean | Difference in Mean | SSD | Difference in SSD |
|---|---|---|---|---|
| 1 (Reference) | 121.12 | 29 | 67.04 | 0 |
| 1 (Summarizer) | 92.38 | | 67.15 | |
| 2 (Reference) | 52.24 | 32 | 25.24 | 17 |
| 2 (Summarizer) | 84.39 | | 42.90 | |
| 3 (Reference) | 71.17 | 7 | 39.30 | 5 |
| 3 (Summarizer) | 78.97 | | 44.47 | |

The Language Technologies Research Centre (LTRC), IIIT [19] provides an online generic summarizer. We fed the three web documents to the summarizer and its performance was 37 percent. We also used SweSum and Pertinence [20] with the three documents and their performances were less than 40 percent in summarizing the documents. The performance evaluation of these summarizers is depicted in Fig. 6.

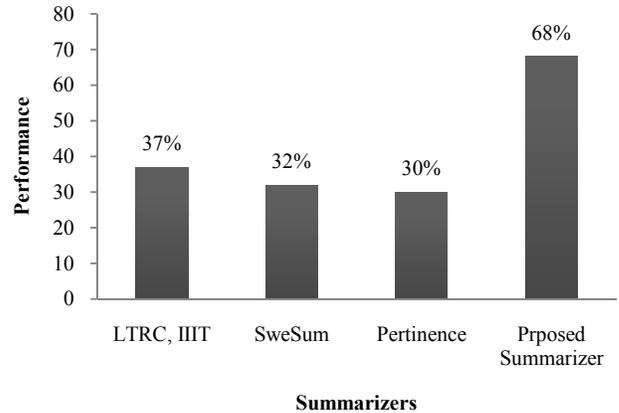

Figure 6.  Performance evaluations of web document summarizers

## V. CONCLUSIONS

In this paper, we proposed an approach to summarize domain-specific text from single document using linguistic and statistical methods and a representative corpus as a reference. The novelty of this approach is to rank sentences based on sentence and subject weight and to extract sentences from web documents measuring their relevance with the textual information in the corpus. We compared the means of web documents and the mean of the corpus to choose three representative web documents and summarized them with the proposed summarizer. 16 different human subjects also produced summaries of the documents and we produced a reference summary from them. We compared these summaries and showed that our proposed summarizer performs better than other web document summarizers.